\documentstyle[11pt,newpasp,epsf]{article}
\def\hi{{\sc H$\,$i}}
\def\et{et al.}
\begin{document}

\title{Mergers of Galaxies from an HI Perspective}
\author{J. E. Hibbard}
\affil{NRAO, 520 Edgemont Road, Charlottesville VA 22903}

\begin{abstract}
\hi\  spectral line mapping studies are a unique probe of morphologically
peculiar systems. Not only does HI imaging often reveal peculiarities
that are totally unsuspected at optical wavelengths, but it opens a
kinematic window into the outer dynamics of these systems that is
unmatched at any other wavelength.  In this review I attempt to
summarize what we have learned from such studies, and what we may hope
to learn from them in the future.
\end{abstract}


\section{Introduction}

\begin{figure}
\plotfiddle{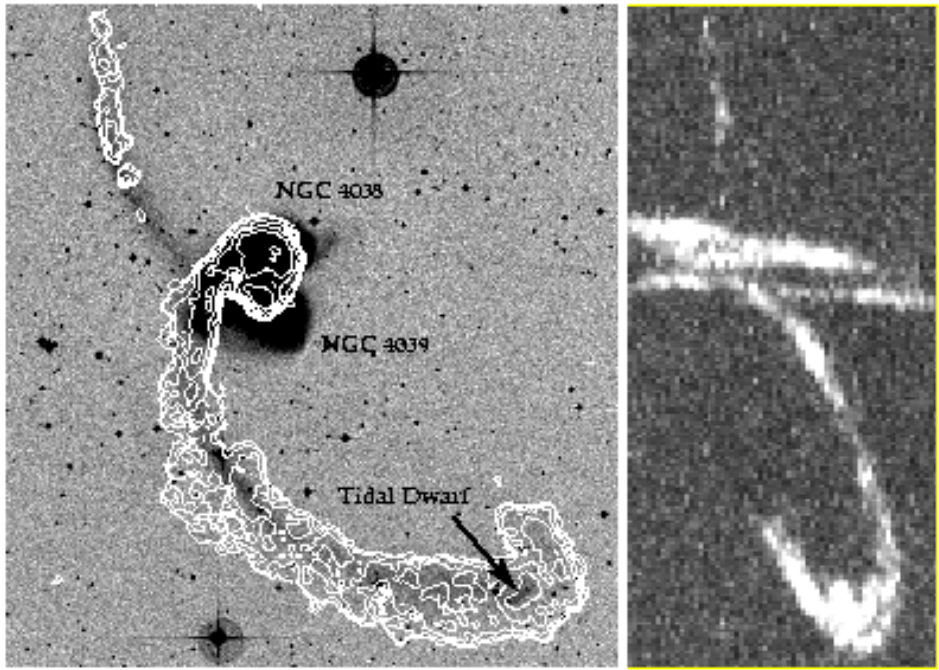}{3.1in}{0}{100}{100}{-180}{-10}
\caption{\scriptsize VLA C+D array \hi\  observations of NGC 4038/9
(Hibbard, van der Hulst \& Barnes, in preparation).
Left:  In this and all subsequent figures contours indicate 
the integrated \hi\  emission and greyscales the optical morphology. 
Right: \hi\  position-velocity plot, with declination along the 
$y$-axis and line-of-sight velocity along the $x$-axis.} 
\label{n4038}
\end{figure}

Spiral galaxies, particularly later types, tend to be rich in neutral
hydrogen.  Much of this gas is found in the outermost regions of the
disks, which are the first regions to be perturbed during tidal
interactions.  As such, the structure of the gas-rich material thrown
off in such encounters will bear the spatial and kinematic imprint of
the encounter dynamics.  Mapping the distribution and line-of-sight
velocity of the atomic gas in the 21cm line of neutral hydrogen (\hi)
is therefore a unique and powerful tool for investigating these
violent events.  As an example, Figure~\ref{n4038} shows \hi\
observations of the classical on-going disk-disk merger NGC 4038/9,
``The Antennae''.  This figure emphasizes the kinematic and spatial
continuity of tidal features.  It is this continuity that makes \hi\
observations so powerful for investigating on-going mergers and their
evolved remnants.  The tails are generally much too faint to map the
stellar kinematics, and ionized emission tends to be confined to a few
localized regions of star formation.  \hi\ mapping is very often the
only way to obtain such information.

At present, at least 140 on-going interactions, mergers, or merger
remnants have been mapped in the 21cm line of neutral hydrogen.  This
includes such classes of objects as interacting doubles, major mergers,
evolved merger remnants, shell galaxies, ring galaxies, polar ring
galaxies, compact groups, and ellipticals with extended \hi\ debris.  In
the remainder of this review I will highlight some of what we have
learned from these observations.  It is beyond the scope of this
review to summarize the wealth of knowledge obtained on each of the
more than 140 systems observed, and I will instead highlight a few
global themes. For additional details, the reader is
directed to the proceedings edited by Arnaboldi \et\ (1997), especially
the contributions by van Gorkom \& Schiminovich, Morganti \et,
Schiminovich \et, and Oosterloo \& Iovino.  See also the recent \hi\
reviews of mergers by Sancisi (1997), of compact groups by
Verdes-Montenegro \et\ (1999), of ring galaxies by Appleton \&
Struck-Marcell (1996), and of polar ring galaxies by Sparke (these
proceedings). 

\section{True Fraction of Peculiar Galaxies}

\hi\  mapping very often reveals a markedly different dynamical picture
of systems than suggested by the distribution of the optical light.
Particularly striking examples are: the extensive tidal streamers
found connecting the members of the M81 group (van der Hulst 1979, Yun
\et\  1994); the 200 kpc rotating \hi\  ring in the M96 group (Schneider
\et\  1989); a pair of purely gaseous tidal tails emerging from the E4
galaxy NGC 1052 (van Gorkom \et\ 1986), from the E2 galaxy NGC 5903
(Appleton, Pedlar \& Wilkinson 1990) and from the Sa galaxy NGC 7213
(Hameed, Blank \& Young in preparation); the \hi\ bridge/tail
morphology of the ``Virgo Cloud'' \hi\ 1225+01 (Giovanneli \et\ 1991,
Chengalur \et\ 1995); plumes of \hi\ pulled off the Sb galaxy NGC 678
by the Epec galaxy NGC 680 and the associated intergalactic \hi\ cloud
(van Morsel 1988).  As a result of these and many similar discoveries,
we conclude that the true fraction of peculiar objects must be
considerably larger than derived from purely optical studies.  Based
on \hi\ studies, Sancisi (1997) suggest that at least one in four
galaxies has suffered a recent merger or experienced an accretion
event.

\begin{figure}
\plotone{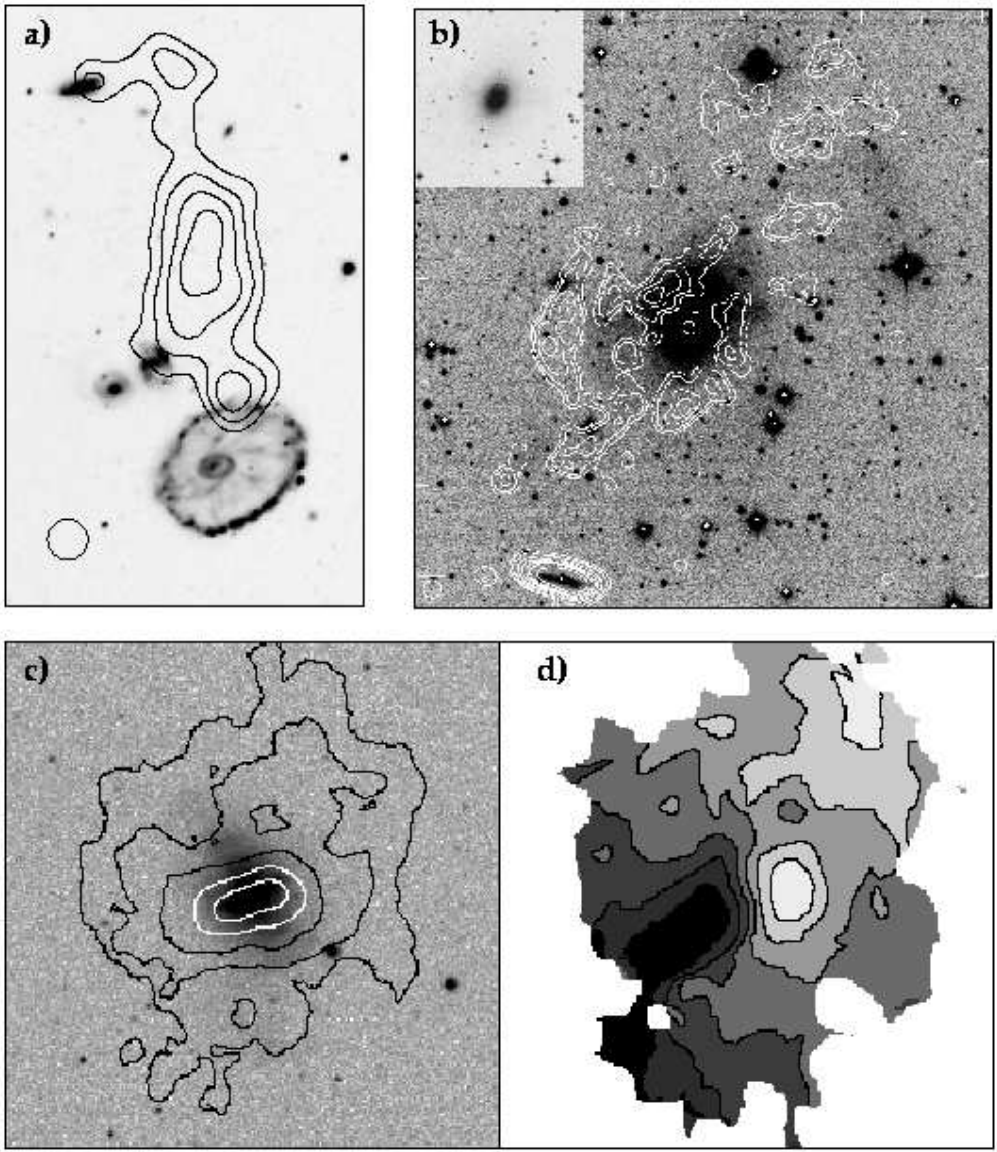}
\caption{(a) VLA D-array observations of the 
Cartwheel Ring Galaxy (Higdon 1996). The \hi\  plume reveals 
a connection between the ring and the northernmost galaxy. Prior 
to this observation it was not clear which of the three galaxies 
in the field collided with the ring galaxy.
(b) VLA C+D-array observations of the shell elliptical NGC 2865
(Schiminovich \et\  1995). Inset shows the main body of the galaxy. 
The \hi\ has rotational kinematics.
(c) \& (d) VLA B+C array observations of the proto-typical shell 
galaxy Arp 230 (Schiminovich, van Gorkom \& van der Hulst, in 
preparation). Left: integrated \hi. Right: \hi\  velocity field, 
showing that the outer \hi\  is arranged into a rotating disk.}
\label{shells}
\end{figure}

Even in systems already identified as optically peculiar, \hi\  mapping
frequently reveals structures that provide critical insights into
their dynamical nature by revealing connections not seen at other
wavelengths (e.g., Figure \ref{shells}).  Examples include:
tidal \hi\  in QSOs (Lim \& Ho 1999); the nearly 200 kpc long tidal
plumes emerging from the ring galaxy Arp 143 (Appleton \et\  1987; see
Figure \ref{a143}) and the IR luminous starburst Arp 299 (Hibbard \&
Yun 1999); the 275 kpc diameter \hi\  disk around the mildly
interacting system Mrk 348 (see Figure 3); the \hi\  tail and
counter-arm in the starburst galaxy NGC 2782 (Smith 1991); the extended tidal
streamers in the starburst/blowout system NGC 4631 (Weliachew \et\
1978); the extended disk and streamers in the dIrr NGC 4449 (Hunter
\et\  1998); two \hi\  tails emerging from the blue compact dwarf II Zw
40 (van Zee \et\ 1998).  The fact that these features are easily
visible in \hi\  but lack optical counterparts is likely due to the
fact that in disk galaxies \hi\  is generally more extended than the 
stars.

It is not clear if \hi\  mapping is the most efficient means for
revealing low-level peculiarities.  When a similar amount of observing 
time ($\sim$ few to a dozen hours) is invested in deep optical imaging,
some remarkable results have emerged: faint optical loops
and streamers have been discovered around what were long thought to be
normal unperturbed disk galaxies (see Malin \& Hadley 1997, Zheng
\et\  1999).  While the optical observations do not include the
kinematic information provided by \hi\  observations, they may be the
only signatures of very evolved interactions, when the \hi\  has faded 
away or been ionized.

\section{Global Dynamics of Merging Systems}

As demonstrated by Toomre \& Toomre (1972) (and re-affirmed many times
since, e.g.  Barnes 1998), tidal features develop kinematically.  As a
result, they have a simple kinematic structure, with energy and angular
momentum increasing monotonically with distance along the tail (Hibbard
\& Mihos 1995).  Because of this simple kinematic structure, \hi\
observations provide a uniquely useful constraint on N-body simulations
of gas-rich mergers (e.g.  Combes 1978, Combes \et\  1988, Hibbard \&
Mihos 1995, Yun 1997, Barnes 1998).  While the primary parameters that are
fit in this exercise are the physically uninteresting angles describing
the orientations of the disks and the viewing perspective, the model
matching gives us the confidence to explore the evolutionary history of
mergers beyond the best-fit time. By running the simulations forward in 
time, we can explore the late-stage merger evolution for clues on the 
expected morphology of the remnants and the distribution of material at 
large radii in the halos around the remnants.  

Because much of the tidal material remains bound to the remnant, it
will eventually reach an apocenter, turn around, and move back inwards
in the potential.  There will therefore be a constant rain of tidal
material back onto the remnant.  Material which falls back while
the potential is still violently relaxing will scatter and be mixed
throughout the remnant body.  Material which returns after the
potential has relaxed will wrap coherently, forming shells, loops and
other ``fine structures" (Hernquist \& Spergel 1992).  Because of its
high energy and angular momentum, the material which falls back later
will fall back to larger and larger radii, forming loops rather than
shells. At late times, the material outside of the loops will have a
low density and may be ionized by the intergalactic UV field (Corbelli
\& Salpeter 1993, Maloney 1993) or the remnant itself (Hibbard, Vacca
\& Yun 1999).  We would therefore expect evolved disk merger remnants
to exhibit partial rings of \hi\ with a rotational signature (since
the loops correspond to turning points where the radial velocity goes
to zero), lying outside the remnant body.  This is exactly what has
been found around a number of shell galaxies (Schiminovich \et\
1995; Fig.~\ref{shells}b--d).

Meanwhile, the loosely bound tidal material in the outer regions
continues to travel outward.  This material has radial periods $\sim$
many Gyr and azimuthal periods even longer than this (Hibbard 1995).
As a result, the tidal material will not give rise to a smooth,
spherical halo of material; instead there will be specific regions of
higher column density material with a low filling factor extending to
very large radii.  At late times, the atomic gas will be too diffuse
to be detected in emission, and may anyways be largely ionized by the
intergalactic UV field.  Therefore the tidal features mapped in \hi\
are likely the denser neutral peaks of a more extended distribution.
This material should be detectable in absorption against background
sources (Carilli \& van Gorkom 1992).

\section{Galaxy Transformation: Spirals to Ellipticals}

The evidence that at least some mergers of gas-rich disk galaxies can
make elliptical-like remnants is very strong (e.g.~Schweizer 1998). 
Whether these merger remnants are true ellipticals or anomalous in
some manner is still
a matter of debate (see van den Marel \& Zurek, these proceedings). 
\hi\  observations addressed one important aspect of this question: do
mergers get rid of the atomic gas of the progenitors? It has often
been stated that \hi\ will be ejected into the tidal features, but in
fact at least as much (and likely much more) outer gas should be sent
into the inner regions as is found in the tidal tails (see Fig.~15 of
Toomre \& Toomre 1972).  It was therefore reassuring to find that
progressively more advanced merging systems have less and less atomic
gas in the bodies of the remnants (Hibbard \& van Gorkom 1996).

It was not clear how most of the original atomic gas was removed from
the inner regions, or how they remain largely \hi\ free in light of
the \hi\ which continues to fall back from the tidal regions. Recent
observations have shed some light on this subject, by showing that two
processes --- galactic superwinds and ionization by continued star
formation --- can have a strong effect on the observability of tidal
\hi\ (Hibbard, Vacca \& Yun 1999). Superwinds are likely to be
important in helping the most gas-rich systems get rid of much of
their cold gas reservoirs, but the wind phase is short lived, and
would not explain the continued removal of returning tidal \hi. Simple
calculations suggest that the UV flux from on-going starformation is
sufficient to ionize diffuse \hi\ in the tidal regions (see also
Bland-Hawthorn \& Maloney 1999).

Photoionization is an attractive mechanize for explaining tidal
features which are gas rich in the outer radii, but gas poor at
smaller radii (e.g., the northern tail of The Antenna,
Fig.~\ref{n4038}; Arp 105 Duc \et\ 1997; NGC 7252 Hibbard \et\ 1994).
The dynamics of tail formation require that the gas-rich outer radii
of the progenitor disks extend all the way back into the remnant (see
Fig.~2 of Toomre \& Toomre 1972).  The geometry of a preliminary
numerical fit to the NGC 40389/9 data (Hibbard, van der Hulst \&
Barnes, in preparation) suggest that the northern tail has an
unobstructed sightline to the numerous starforming regions in the disk
of NGC 4038, while the southern tail does not, explaining why it
remains gas rich along its entire length. This process may explain how
merger remnants remain gas poor in the presence of the continued
return of tidal \hi. Such an on-going process is required if remnants
are to evolve into normal ellipticals in terms of their atomic gas
content.

\section{Galaxy Transformation: Other Beasts}

\begin{figure}
\plotone{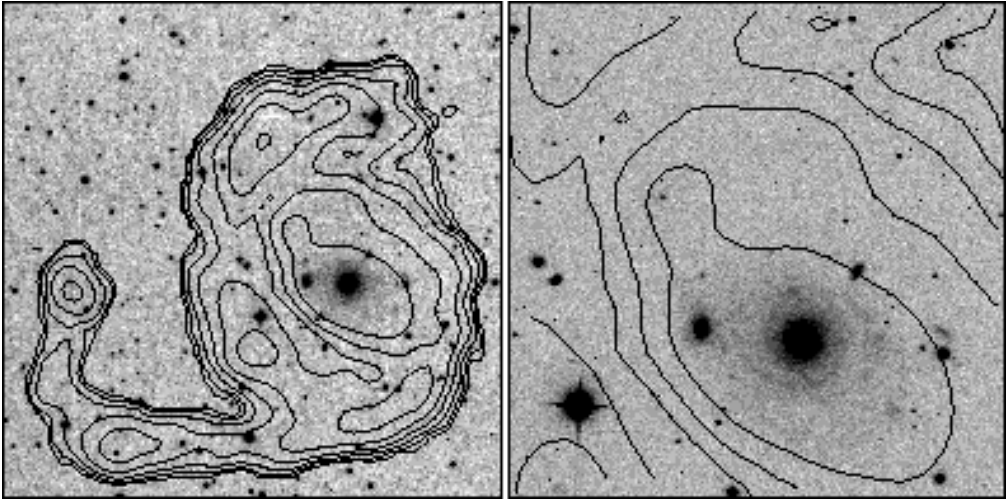}
\caption{VLA D-array observations of Mrk 348
(Simkin \et\  1986). Left:  Full field of view (image is 300 kpc on 
a side). Right: close-up of inner regions and companion (image is 
100 kpc on a side).}
\end{figure}

Is the ultimate evolutionary product of disk-disk mergers an
elliptical with fine structure? Here again \hi\ observations provided
evidence for unexpected merger products.  In particular, a number of
on-going mergers and merger remnants are found to have large gaseous
disks with rotational kinematics.  Particularly good examples are Arp
230 (Fig.~\ref{shells}c\&d), NGC 520 (Hibbard \& van Gorkom 1996), and
MCG -5-7-1 (Schiminovich, van Gorkom \& van der Hulst in preparation).
The very faint loops and streamers imaged around normal disk galaxies
(Malin \& Haley 1997, Zheng \et\  1999) support the idea that some 
disk systems may have had a violent origin or experienced a major 
accretion event. 

Finally, there are some systems which simply do not seem to conform to
the standard interaction picture.  One such example is Mrk 348
(Fig.~3).  The main difficulty with the tidal interpretation for this
system is that the scale of the \hi\ is tremendous (diameter $\sim$280
kpc), and two thirds of the neutral hydrogen ($1.4\times 10^{10}
M_\odot$ out of a total of $2.1\times 10^{10} M_\odot$) lies outside
of the highest contour in Fig.~3, i.e., outside the region containing
both the companion and all of the optical light of the disk.  It
simply does not seem possible that the small companion seen in Fig.~3b
could have raised this much material to such large radii.  It may be
that the progenitor was a very gas-rich low surface brightness galaxy
like Malin 1 (Impey \& Bothun 1989, Pickering \et\ 1997).  A more
intriguing possibility is that the neutral gas may have condensed out
of an extensive halo of ionized gas.  In this regard it is interesting
to consider the NGC 4532/DDO 137 system, which has a very irregular
distribution of \hi\ lying mostly outside of the optical galaxies.
Hoffman \et\ (1999) suggest that the \hi\ clumps are simply neutral
peaks in sea of mostly ionized hydrogen.  The existence of such a sea
of baryons may mean that full scale galaxy formation continues to the
current epoch.

\section{Galaxy Formation: Tidal Dwarf Galaxies}

\begin{figure}
\vbox {
  \begin{minipage}[l]{0.37\textwidth}
    {\centering \leavevmode \epsfxsize=\textwidth 
      \epsfbox{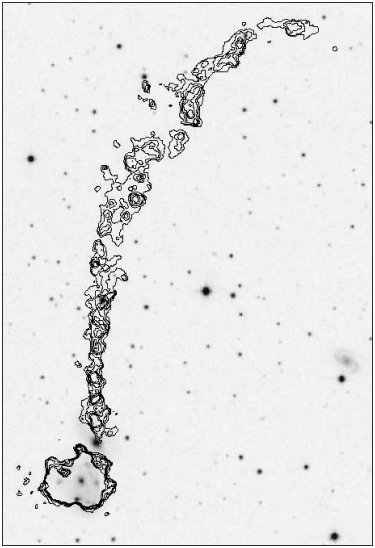}}
  \end{minipage} \  \hfill \
  \begin{minipage}[r]{0.63\textwidth}
\caption{VLA B+C+D-array observations of the Ring Galaxy Arp 143 
(Higdon \& Appleton, in preparation; see also Appleton et al.~1987). 
This purely gaseous tidal tail contains a wealth of structure on 
5$^{\prime\prime}$ scales, with peaks up to $10^{21} {\rm cm}^{-2}$. 
The present observations do not have the velocity resolution to 
address whether individual clumps are self
gravitating.}
  \end{minipage}
}
\label{a143}
\end{figure}

As the name implies, ``tidal dwarf galaxies'' are concentrations of
stars entrained within tidal tails and believed to be gravitationally
bound (Schweizer 1978). These systems have received considerable
observational attention recently (e.g. Duc 1995, Hunsberger \et\
1996). However, because the inter-clump tidal material is so faint,
\hi\  mapping studies provide the {\it only} means of determining
whether the local luminosity enhancements are kinematically distinct
from the surrounding material, and this has only been done for a few
systems (Hibbard \et\  1994; Hibbard \& van Gorkom 1996; Duc \et\
1997).

Within tidal tails there is a wealth of substructure on many scales.  It
ranges from small, dense gaseous knots within purely gaseous features
(e.g., Figure \ref{a143}), to small luminosity enhancements within
optical tidal tails (e.g.~Hutchings 1996, Hunsberger \et\ 1996), to
dwarf-sized condensations of gas and stars fully embedded within a tidal
tail (e.g.~NGC 7252, NGC 3921 Hibbard \& van Gorkom 1996; NGC 4038/9
Fig.~\ref{n4038}a); and finally to separate (and often separately
classified) optical dwarfs entrained within mostly gaseous tidal
features (e.g., M81/NGC 3077, van der Hulst 1979; NGC 4027, Phookun \et\
1992; NGC 520/UGC 957, Hibbard \& van Gorkom 1996; Arp 105, Duc \et\ 1997;
NGC 5291, Malphrus \et\ 1997).  An outstanding question is whether there
is an evolutionary link between any/all of these categories of structures. 

\section{Timing of Starbursts}

The 100 kpc scale tidal features imaged in \hi\ emanating from
starbursting systems suggest that the interaction and starburst
timescales are quite different.  For example, the starburst in the IR
luminous merger Arp 299 has an age of $<$30 Myr while the 180 kpc tail
was launched about 700 Myr ago (Hibbard \& Yun 1999).  Similarly, the
tails of the Antennae (Fig.~\ref{n4038}) suggest an interaction
timescale of $\sim$ 500 Myr.  This object has a population of
star-clusters with an age of $\sim$ 500 Myr as well as a population
that is currently forming massive stars (Whitmore \& Schweizer 1995).
These observations suggest that interaction induced starbursts are not
isolated to either first periapse (when the tails are launched) or the
final merger, but rather are episodic (cf Noguchi 1991).  While the
closeness of the nuclei of the ultraluminous IR mergers suggests that
the most intense starbursts occur when the progenitor nuclei are
coalescing, it does not necessarily follow that the bulk of the stars
are formed during this short-lived phase.

This fact has important repercussions for the expected observational
characteristics of merger remnants.  If most of the
interaction-induced starformation takes place at the moment of final
coalescence, the burst population is expected to be confined to the
inner few 100 pc of the remnant, leaving an anomalously bright central
core (Mihos \& Hernquist 1994) with the characteristics of a younger,
more metal enriched population.  The lack of such signatures in shell
galaxies is taken to mean that they could not have formed via major
mergers (Silva \& Bothun 1998).  However, if much of the
post-interaction population formed over the whole $\sim$ 1 Gyr
timescale of the merger, then the ``burst" population will be spread
more widely through the remnant, leaving much more subtle
observational signatures.  

\section{Conclusion}

\begin{figure}
\plotfiddle{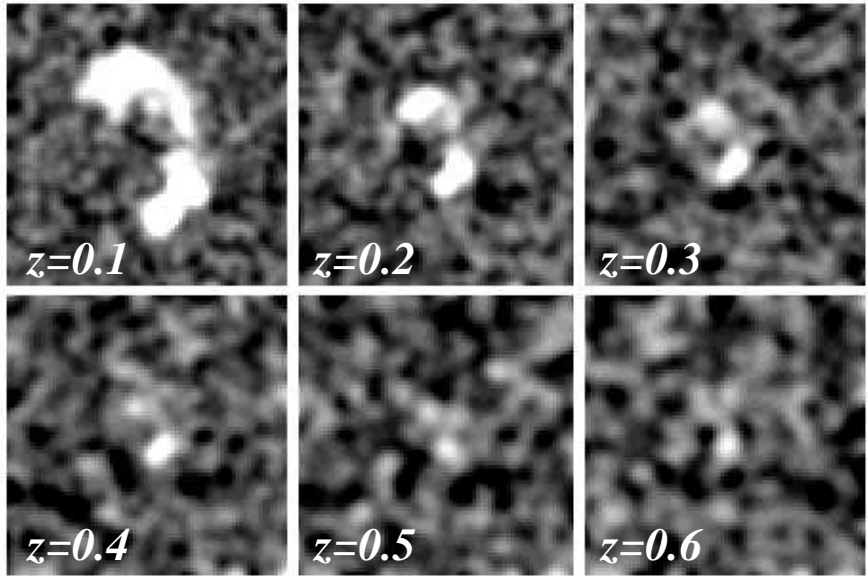}{2.75in}{0}{80}{80}{-165}{-10}
\caption{\scriptsize 
Simulated deep HI image with expanded VLA: 
The IR luminous merger Arp 299 (Hibbard \& Yun 1999), 
as viewed at redshifts from $z=$0.1 to 0.6 ($H_o$=75, $q_o=$0.1).}
\label{upgrade}
\end{figure}

\hi\  spectral line mapping is a powerful diagnostic tool for
investigating interacting and peculiar galaxies.  In concert with
numerical simulations, such observations provide insight into the
transformation and formation of galaxies, the distribution of material
in the halos of galaxies, the timing of interaction-induced
starbursts, and the possible evolutionary products of mergers.

An important outstanding questions is whether many normal systems
formed via mergers.  While a merger origin for most galaxies is a
generic result of hierarchical structure formation scenarios, there
are continued claims that merger remnants will differ from normal
ellipticals (Mihos \& Hernquist 1994, van den Marel \& Zurek, these
proceedings).  \hi\ observations can help address this question by
identify evolved remnants of gas-rich mergers via the amounts
and structure of any remaining tidal \hi. Once identified, the
structure of these remnants should be compared to ellipticals. If they
are indeed different, then this might mean that the Hubble Sequence
evolves with redshift, such that the merger of present day spirals
evolve into ellipticals with different characteristics than
present day ellipticals, and conversely that present day ellipticals
had progenitors which differed in some manner from present day disk 
galaxies.

With future cm wave facilities we should be able to address the
cosmological aspect of this question.  For instance, an expanded VLA
(cooled low frequency receivers, greatly expanded correlator) will be
able to detect \hi\ out to redshifts $\sim$1.  We should be able to
image the gas-rich tidal features out to redshifts of $z\sim$0.5
(Figure \ref{upgrade}). We will thus be able to constrain the number
density of gas-rich mergers at these redshifts, which will tell us how
large the population of gas-rich merger remnants should be at the
present epoch. 

\acknowledgments

I thank Jacqueline van Gorkom for useful discussions and a careful
reading of this manuscript, Jim Higdon \& Min Yun for providing
figures for my talk, and the organizers for the opportunity to attend
such an interesting meeting.

\end{document}